# The Art of the Meta Stream Protocol

## Torrents of Streams

## Christophe De Troyer[a], Jens Nicolay[a], and Wolfgang De Meuter[a]

a   Vrije Universiteit Brussel, Brussels, Belgium

**Abstract**     **Context** The rise of streaming libraries such as Akka Stream, Reactive Extensions, and LINQ popularized the declarative functional style of data processing. The stream paradigm offers concise syntax to write down processing pipelines to consume the vast amounts of real-time data available today.

**Inquiry** These libraries offer the programmer a domain specific language (DSL) embedded in the host language to describe data streams. These libraries however, all suffer from extensibility issues. The semantics of a stream is hard-coded into the DSL language and cannot be changed by the user of the library.

**Approach** We introduce an approach to modify the semantics of a streaming library by means of meta-programming at both run-time and compile-time, and showcase its generality.

**Knowledge** We show that the expressiveness of the meta-facilities is strong enough to enable push and pull semantics, error handling, parallelism, and operator fusion.

**Grounding** We evaluate our work by implementing the identified shortcomings in terms of a novel stream meta-architecture and show that its design and architecture adhere to the design principles of a meta-level architecture.

**Importance** The state of the art offers plenty of choice to programmers regarding reactive stream processing libraries. Expressing reactive systems is otherwise difficult to do in general purpose languages. Extensibility and fine-tuning should be possible in these libraries to ensure a broad variety of applications can be expressed within this single DSL.

## The Art, Science, and Engineering of Programming

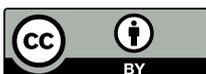

## 1 Introduction

Stream programming is used for expressing computations that are naturally specified as a functional pipeline of operators between data sources and sinks. Examples of this include GUI programming [8, 15, 25], distributed event-driven systems such as IoT applications [9, 14], and complex event processing [27]. Because of its many applications, stream programming has become omnipresent in contemporary software [1, 9, 12] and most mainstream programming languages offer stream processing libraries or embed a domain-specific language (DSL) for this purpose. Because the concepts and basic building blocks are quite homogeneous across most stream implementations, all implementations can in fact be considered as a DSL for the stream programming domain, and we will therefore use the term *stream DSL* to refer to these implementations.

Contemporary and widely used stream DSLs such as Akka Streams [13], RxJava [18], and GenStage [7] offer a high-level flow-based API to construct and manipulate streams of data by using functional-inspired building blocks called "operators", such as map and filter. These basic building blocks can be classified into two categories, which are central to our problem statement:

- *Functional operators* such as map, filter, and scan, that manipulate the data flowing through streams.
- *Non-functional operators* that manipulate *how* the data streams through the stream such as buffer and async.

As we already mentioned, this API for expressing the *functional* requirements of a stream program are homogeneous between the different DSLs. It is a different story, however, when it comes to expressing the *non-functional* requirements of stream programs. Surveying four existing stream DSLs embedded in popular general-purpose languages [21] (Akka Streams in Scala, Java Streams and RxJava in Java, and GenStage in Elixir), we identified three problems with respect to expressing non-functional requirements in these DSLs. These issues have been addressed in the past in the context of aspect-oriented programming [10, 11] and we wish to address a subset of them in our work.

1. Stream DSLs provide non-functional extensions in a non-canonical and ad-hoc fashion, hindering transplantability of programs between languages.
2. Entanglement of functional stream logic and non-functional stream execution logic, leading to reduced readability and maintainability of stream programs.
3. Limited or no means for extending and adapting the stream execution semantics from within the DSL beyond what is built into language.

In this paper, we propose meta-programming as an approach to overcome the problems we identified in a structured manner. More in particular, we examine how meta-programming can be used for expressing non-functional requirements in stream programming in a language agnostic way, disentangled from the functional requirements, and extensible by stream programmers. Central to realizing our approach is the meta-level architecture $\mu$CREEK. Our approach is two-pronged: run-time meta-programming ($\mu$CREEK$_r$) and compile-time meta-programming ($\mu$CREEK$_c$). $\mu$CREEK$_r$ allows the programmer to modify the run-time behaviour of a stream (e.g., push vs. pull based). $\mu$CREEK$_c$ hands the programmer control over the construction of streams at compile-time (e.g., introduce parallelism or timestamping). To define and implement $\mu$CREEK, we designed CREEK, an extensible stream DSL that incorporates our meta-programming facilities. To validate our approach we implement 5 examples prominent in related work [2, 20]: operator fusion, push and pull semantics, error handling, timestamping, and operator-level parallelism.

The contributions of the work are as follows:

- CREEK, an extensible stream DSL with stream-based meta-programming facilities.
- $\mu$CREEK, a stream meta protocol that offers the programmer control over the structure of the streams ($\mu$CREEK$_c$) by means of DAG rewriting, and the run-time semantics of the underlying stream DSL ($\mu$CREEK$_r$) by exposing and hooking into a high-level communication protocol, all in the same paradigm (i.e., streams) as the base language.
- An evaluation of the expressivity and applicability of $\mu$CREEK by expressing prominent problems from related work in $\mu$CREEK.

**Overview**  In Section 2 we introduce and discuss a number of examples to demonstrate the three problems we identified, and use them to motivate our approach. To explain our approach we first introduce CREEK (Section 3), our extensible stream DSL, and present its compile-time meta-protocol $\mu$CREEK$_c$ (Section 4), and conclude with its run-time meta-protocol, $\mu$CREEK$_r$ (Section 5). We validate our meta-protocols by showing that $\mu$CREEK$_c$ and $\mu$CREEK$_r$ can express common shortcomings of stream DSLs identified in related work such as operator fusion, operator parallelism, and alternative propagation algorithms (e.g., push vs. pull) without making changes to the base language. We evaluate the performance overhead of $\mu$CREEK$_r$ in CREEK in Section 6. After discussing related work (Section 7), we conclude the paper and mention some directions for future work and avenues for future research. (Section 8).

## 2   Problem Statement

We have presented three problems in the state of the art regarding stream DSLs in contemporary programming languages in Section 1. In this section we substantiate these claims. More specifically, we show by example that contemporary stream DSLs provide non-functional extensions in a non-canonical and ad-hoc fashion (Section 2.1), that they entangle functional and non-functional stream logic (Section 2.2), and that there are limited or no means for extending and adapting the stream execution semantics from within the DSL beyond what is offered by the language (Section 2.3).

Although we will discuss each problem in the context of only one or two of the prominent streaming libraries we surveyed for this purpose, we observed all languages suffer from at least one or more of the posited problems.

■ **Listing 1** Sequential program containing a long-lasting computation (left), and its parallel version (right) in RxJava.

```
1  source
2  .map(i ->
3    longLastingComputation(i))
4  .subscribe()
```

```
1  source
2  .subscribeOn(Schedulers.computation())
3  .flatMap(val ->
4      just(val)
5      .subscribeOn(Schedulers.computation())
6      .map(i -> longLastingComputation(i))
7      .subscribeOn(Schedulers.single()))
8  .subscribe();
```

■ **Listing 2** Sequential program (left) from Listing 1 and its parallel version (right) expressed in Akka Streams.

```
1  source
2  .map((i) =>
3    longLastingComputation(i))
4  .runForeach(println)
```

```
1  val processor: Flow[Int, Int, NotUsed] =
2    Flow.fromGraph(GraphDSL.create() { implicit b =>
3      val balance = b.add(Balance[Int](2))
4      val merge = b.add(Merge[Int](2))
5      val f = Flow[Int].map(longLastingComputation)
6
7      balance.out(0) ~> f.async ~> merge.in(0)
8      balance.out(1) ~> f.async ~> merge.in(1)
9      FlowShape(balance.in, merge.out)
10   })
11   source.via(processor).runForeach(println)
```

## 2.1  Problem 1: Lack of Canonical Non-functional Operators

Stream execution semantics is concerned with how a definition of a stream is executed at run-time to reduce the stream to a result. Given the wide range of applications in which stream DSLs are applicable, there are situations in which the default stream execution semantics offered by a stream DSL does not (or no longer) match with the intended semantics. Stream DSLs cater to this problem by offering non-functional operators to manipulate how a stream is executed with respect to propagation semantics (push, pull, back-pressure, . . . ), concurrency and parallelism, buffering, error handling, etc. These operators are used to improve on one or more non-functional requirements such as scalability, performance, monitoring, maintainability, or recoverability. However, while the typical building blocks for expressing the functionality of stream programs are extremely similar across different stream DSLs, this is not the case for the non-functional operators. Moreover, different DSLs offer different sets of non-functional operators for supporting different non-functional requirements.

### 2.1.1  Example: Ad-hoc Parallelization in RxJava and Akka Streams
Suppose that a stream program must map some long-lasting computation (represented by the function longLastingComputation) over a stream of values, and that applying this operator in parallel would increase throughput. Listing 1 shows a small RxJava program containing the `longLastingComputation`. The left-hand side is the sequential version, and the right-hand side is a parellelized version. Due to the poor integration of parallelism in RxJava, the stream programmer is forced to modify the layout of the stream substantially to facilitate for the parallelism (lines 3–7), making its functionality less readable. Listing 2 contains the same sequential example program on the left, and the parallelized version on the right, but written in Akka Stream. As in RxJava, parallelizing a single operator introduces a significant amount of non-essential complexity, significantly increasing the number of LoC. From the code in Listings 1 and 2 we make the following observations:

- Although both RxJava and Akka Streams allow the parallelization of operators, the mechanism to do so is very different. RxJava requires the programmer to manually place computations on different threads of computation, while Akka Stream requires the programmer to manually create parallel pipelines of computation.
- Although parallelization is a non-functional concern, programmers in RxJava and Akka Streams have to modify the original stream to express this requirement.

### 2.1.2 Example: Error Handling in Akka Streams and RxJava

■ **Listing 3**  Error handling in Akka Streams using Decider. All IllegalArgumentExceptions cause a restart, and other exceptions stop the stream execution.

```scala
 1 val decider: Supervision.Decider = {
 2   case _: IllegalArgumentException => Supervision.Restart
 3   case _                           => Supervision.Stop
 4 }
 5 val flow = Flow[Int]
 6   .scan(0) { (acc, elem) =>
 7     if (elem < 0) throw new IllegalArgumentException("negative not allowed")
 8     else acc + elem
 9   }
10   .withAttributes(ActorAttributes.supervisionStrategy(decider))
11 val source = Source(List(1, 3, -1, 5, 7)).via(flow)
12 val result = source.runWith(Sink.seq)
```

Stream DSLs offer facilities for the programmer to deal with exceptions during stream execution. For instance, Akka Streams modularizes exception handling of a stream by extracting the behaviour into a separate object called a `Decider`. The programmer defines what should happen in case of a specific exception class. The example in Listing 3 defines a `Decider` object that, when an `IllegalArgumentException` occurs, restarts the stream entirely. If the stream produces any other type of error the entire stream is aborted and is reduced to an error value. If the source of the stream produces the erroneous value again the stream will restart, possibly creating an infinite loop if no limit is placed on the restarting.

Decider behavior can be added in a modular fashion, separated from the actual stream logic. For instance, the stream program in Listing 3 instructs the `scan` operator to employ the `decider` object. However, not all stream DSLs offer this kind of (modular) error handling. In RxJava, for example, one cannot express something similar to the `Decider` object. The only way for the programmer to handle errors is by changing the definition of the stream by inserting additional non-functional operators such as `retry` and `retryWithValue`. We do note that any approach discussed here reexecutes the entire stream. Ignoring an erroneous value is impossible.

### 2.1.3 Problem 1: Conclusion

Stream DSLs feature non-functional operators that address non-functional concerns. However, opposed to the work in Object-Oriented Programming [10, 11], there is no canonical approach to deal with non-functional concerns. Stream DSLs approach non-functional concerns in an ad-hoc fashion, resulting in completely different approaches for the same problem in different DSLs.

## 2.2 Problem 2: Entanglement of Functional and Non-functional Operators

In stream programming the domain logic of the stream (i.e., *what* the stream computes) is a separate concern from the execution semantics of the stream (i.e., *how* the stream executes). The software engineering principle of "Separation of Concerns" dictates that the different concerns of an application should be defined in separate parts of code, separated by a clean interface. This implies that functional and non-functional operators should not be mixed in the specification of a stream. Additionally, using non-functional operators in stream DSLs often forces the specification of the stream program to be modified to suit the inclusion of those operators. In Section 2.1 we have already encountered examples where this was the case; in the larger example that follows we revisit parallelization in Akka Streams.

### 2.2.1 Example: Entangled Parallelization in Akka Streams

Consider the example program in Listing 4, that analyzes a stream of tweets containing the keywords "covid19" and "sars-covid" for its sentiment: positive, neutral, or negative. Every minute, the program prints the ratio of positive, negative, and neutral sentiments to the console. Given that sentiment analysis is a resource-intensive computation, this step is parallelized with the goal of improving the throughput of the entire stream.

When starting out from a completely sequential, program two changes need to be made. Firstly, a parallel pipeline needs to be created that balances the tweets over two sentiment analyzers (lines 5–10 in Listing 4). Secondly, the sentiment analysis step must be factored out into a separate flow (lines 1–3 in Listing 4) so that it can be referenced twice in the parallel pipeline (lines 5–10). While the sequential program is straightforward and declares what should happen to each tweet, the parallel version mixes the domain logic with the parallelization logic.

■ **Listing 4**  Computing the average sentiment of tweets related to COVID-19 in parallel in Akka Streams.

```scala
val analyze = Flow[Status].flatMapConcat((tweet: Status) => {
  var result = Sentiment.computeSentiment(tweet.getText())
  Source(result)
})
val analyzer = Flow.fromGraph(GraphDSL.create() { implicit builder =>
  val dispatchTweets = builder.add(Balance[Status](2))
  val mergeSentiments = builder.add(Merge[(String, Sentiment)](2))
  dispatchTweets.out(0) -> analyze.async -> mergeSentiments.in(0)
  dispatchTweets.out(1) -> analyze.async -> mergeSentiments.in(1)
  FlowShape(dispatchTweets.in, mergeSentiments.out)
})
val source = TwitterActor.mkActor()
val actorRef = source
  .via(analyzer)
  .map(tuple => tuple._2)
  .sliding(60)
  .map((win: Seq[Sentiment]) => {
    (win.count(s => s == Sentiment.POSITIVE),
     win.count(s => s == Sentiment.NEUTRAL),
     win.count(s => s == Sentiment.NEGATIVE))
  })
  .async
  .toMat(Sink.foreach(println))(Keep.left).deploy()
TwitterActor.pipeInto(Array("covid19", "sars-covid"), actorRef)
```

#### 2.2.2  Problem 2: Conclusion

Non-functional concerns are mixed with functional concerns in the same code-base, which burdens the programmer with additional complexity, reduces the readability of the code, and hard-codes the parallel processing factor.

### 2.3  Problem 3: Hard-Coded Execution Semantics

Every Stream DSL offers a specific set of non-functional operators for expressing different non-functional requirements. Because stream DSLs are applicable in many different situations, the intended semantics of the language does not always match the language's semantics. None of the stream DSLs we surveyed, however, provide a structured means for extending and adapting the stream execution semantics from within the DSL itself.

An example of a non-functional requirement is the propagation semantics of a stream. All the stream DSLs we investigated implement either *push-based* or *pull-based* propagation by default, or a variation thereof [6]. Push-based semantics is a good fit for *producer-driven* applications in which, every time the producer produces a new datum, the remainder of the stream computes its result. In this semantics, the producer

■ **Listing 5**  Processing temperature measurements simulating pull-based semantics in Akka
Streams.

```
1  remoteThermometer
2  .buffer(Int.MaxValue, OverflowStrategy.dropHead)
3  .map(f => logToDB(f))
4  .map(f => celsiusToFahrenheit(f))
5  .runWith(foreach(f => {
6    printf("Current temp: %fF\n", f))})
```

dictates the rate at which the stream computes. However, some applications are inherently *consumer-driven* instead: the consumer demands data when it is needed downstream. For consumer-driven applications *pull-based* propagation is a better fit than push-based propagation: the consumer will "pull" data as is needed. The consumer dictates the rate of the stream.

We illustrate this problem in Akka Streams. Akka Streams uses a complex back pressure algorithm [6] that ensures that production and consumption rate in the stream are as compatible as possible. However, in producer-driven scenarios the programmer wants pure push-based semantics. Consider the example of thermometers monitoring a thermal processing pipeline. For safety reasons it may be important that the thermometers measure as frequently as possible, so it is considered a producer-driven stream. Consider the code in Listing 5, that contains an example program that processes a remote stream of measurements from a thermometer. The programmer intended to write a producer-driven stream in this case, but due to the default semantics of the stream library (i.e., back pressured pull-based) this is not the resulting behavior.

### 2.3.1  Example: Emulating Push-Based Propagation in Akka Streams

The stream will run at an equilibrium between the highest rate the thermometer can measure and the rate the system can process the measurements, if the buffer operator is omitted. The buffer operator can process elements at unbounded speed (at the cost of data loss), so when it is inserted in the stream, the equilibrium of the propagation algorithm in Akka Streams between the thermometer and the buffer will be at the maximum rate of the thermometer, resulting in the intended semantics between the thermometer and the buffer, namely push-based semantics.

### 2.3.2  Problem 3: Conclusion

Propagation semantics are hard-coded into the stream DSL implementations. The programmer can only manipulate the streams in a limited fashion by using non-functional operators such as buffer. Moreover, the intricacies of the underlying implementation of the propagation semantics needs to be well understood to use these operators. With the current state of affairs, there is no structured approach to modify the propagation semantics.

■ **Listing 6**  List of the squares of all the even numbers between 1 to 100 written in CREEK.

```
1  defmodule MyDAGs do
2    defdag evens as filter(fn x -> even?(x) end)
3    defdag squares as map(fn x -> x * x end)
4
5    defdag square_of_evens(input, output), do: input -> evens -> squares -> output
6
7    def program() do
8      src = Source.range(1, 100)
9      snk = Sink.all(self())
10     deploy(square_of_evens, [input: src, output: snk])
11     receive do
12       result -> IO.puts "Squares: #{result}"
13     end
14   end
15 end
```

## 3  Stream Programming with CREEK

All of the popular stream DSLs we surveyed suffer from one or more problems related to expressing non-functional requirements that we identified in Section 2. In this paper we argue that meta-programming offers an elegant solution to these problems, but before we introduce our approach in Sections 4 and 5, we introduce CREEK, a prototypical generic stream programming DSL that will serve as a vehicle for explaining our meta-programming approach.

### 3.1  CREEK DSL

We introduce CREEK and its features by means of the example program shown in Listing 6 that computes the list of all the squares of the even numbers between 1 to 100. CREEK is a stream DSL embedded in Elixir [24], a general-purpose language that builds on top of Erlang and runs on the same virtual machine [22]. Every CREEK program must be defined inside an Elixir module by using defmodule. Module MyDAGs in Listing 6 defines three DAGs (Directed Acyclic Graphs) named evens, squares, and square_of_evens using defdag. The former define DAGs that filter out the even numbers from a stream and compute the squares of each number respectively. The latter combines these two into a DAG that creates the squares of all the even numbers. DAGs form the blueprint of a stream in CREEK, and they are defined by combining operators and DAGs into larger DAGs.

A DAG must be deployed with actors that provide the source and sinks. The deploy function is boundary between CREEK and the host language (i.e., Elixir). CREEK offers convenience functions to create source and sink actors, but sources and sinks can be any Elixir actor that implements the Creek protocol (Section 5). In Listing 6 two actors are created using the helper functions Source.range and Sink.all. Source.range(a,b)

■ **Table 1** Functional operators in Creek, which represent a subset of operators that are present in most contemporary streaming DSLs.

| Name | Description |
| --- | --- |
| map(f) | Applies f (arity 1) to each data. |
| filter(f) | Filters out elements for which the predicate f (arity 1) fails. |
| dup(n) | Creates a joint which duplicates the incoming stream n times (i.e., $n$ output ports). ($n > 0$) |
| balance(n) | Creates a joint which alternates the incoming stream over n streams. ($n > 0$) |
| merge(n) | Interleaves n streams into one. ($n > 0$) |
| zip() | Combines two streams into one by creating tuples of each value of each stream. |
| scan(f, acc) | Scans the stream with initial value acc and function f (arity 2). |

creates an actor that will emit all the integers going from a to b. Sink.all(pid) creates an actor that consumes all the values from the stream and sends them as a list to another actor (pid), in the example that is self(), the current process. Because we embed Creek into Elixir, the idiomatic code to handle asynchronous events is message sends. In other languages the idiomatic approach could be futures, promises, or other streams.

### 3.1.1 Operators

Creek uses operators as the basic building blocks for DAGs. An operator is a modular unit of data transformation logic that can be instantiated with an argument. For example, the map operator requires a user-defined function to produce values by transforming input values, and the merge operator manipulates data by combining two streams into one.

Each Creek operator has an input and output arity that determines the number of input and output "ports" they have. For example, the map operator has a single input port and a single output port, while the merge operator has 2 input ports and 1 output port. Creek implements a subset of the functional operators present in contemporary stream DSLs. Table 1 lists the operators in Creek.

Contemporary stream libraries such as Akka Streams [13] and RxJava [18] support higher-order operators (e.g., flatMap and switch) that can work with streams of streams. In essence, these operators rewrite dependencies between operators at run-time. Creek currently does not support higher-order streams. We briefly discuss what the requirements would be on the meta-level to support higher-order operators in Section 5.

### 3.1.2 DAGs

A composition of operators in a DAG forms the blueprint of a stream. The order of the composition determines the order of the transformations that will happen to each datum "flowing" through the DAG. An operator itself is the smallest DAG possible, consisting of only the operator itself. DAGs are created by composing DAGs using the *composition functions* ~> and |||. The vertical composition function (~>) connects two

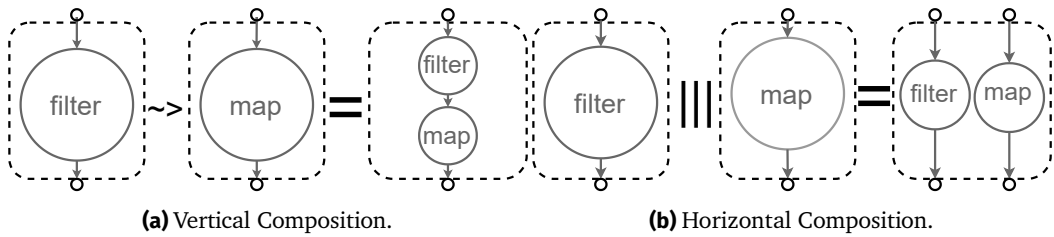

**(a)** Vertical Composition.  **(b)** Horizontal Composition.

■ **Figure 1**  Graphical depiction of the semantics of both composition functions.

DAGs by connecting the output ports of the left DAG to the input ports of the right DAG. When ~> is applied to left operand $a$ with $a_i$ input ports and $a_o$ output ports and right operand $b$ with $b_i$ input ports and $b_o$ output ports where $a_o = b_i$, the result is a DAG with $a_i$ input ports and $b_o$ output ports. For example, the vertical merge of two DAGs with one input and one output port each (Figure 1a) results in a DAG that has one input port and one output port. CREEK requires that $a_o = b_i$; otherwise the program is invalid.

The horizontal composition function (|||) composes two DAGs in parallel. When ||| is applied to DAGs $a$ and $b$ with input and output ports as denoted above, the result is a DAG with $a_i + b_i$ input ports and $a_o + b_o$ output ports. Figure 1b shows the horizontal composition of two DAGs with 1 input and 1 output port each, resulting in a DAG with 2 input ports and 2 output ports.

DAGs in CREEK are classified into two types: *open* and *closed* DAGs. A DAG is called *open* if for any operator in the DAG there is an input or output port that is not connected to another operator. An open DAG is an abstraction mechanism that allows the composition of other open DAGs and individual operators. In CREEK an open DAG is defined with the "defdag $v$ as $d$" syntax.

Conversely, a DAG is called *closed* if it is not open; i.e., every in- and output port is connected to either a DAG, or an "actor socket". An actor socket serves as a placeholder for an actor, for when the DAG is deployed. Listing 6 line 5 defines two actor sockets: input and output. This approach decouples the DAG from actual data sources and sinks when it is deployed, facilitating reuse. In CREEK a closed DAG is defined with the "defdag $v(var*)$, do: $d$" syntax. Each $var$ introduces an actor socket into the scope of the DAG definition, similar to formal parameters in function definitions. Each actor socket can have either one output or input port. In contrast to open DAGs, closed DAGs cannot be composed into larger DAGs.

### 3.1.3  Streams

A stream in CREEK is a *deployment* of a closed DAG in which all actor sockets are replaced by actors that either inject data into the stream, or consume data from the stream. To deploy a closed DAG, a list of actor and label pairs is required where each label in a pair must reference an actor socket in the DAG. Once deployed, a stream runs asynchronously and the only way to interact with the stream is through the sources and sinks given at deployment time. The syntax for deploying a stream in CREEK is deploy($d$, ($actorref$, $var$)*), where $d$ is a DAG, followed by a list of actors and labels.

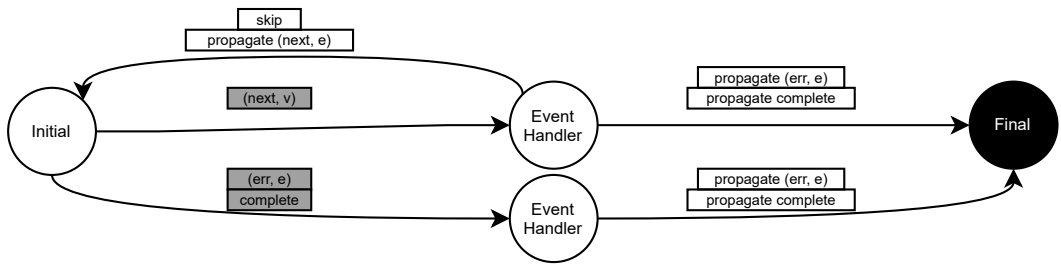

**■ Figure 2** The protocol of an operator represented as an statemachine.

### 3.2 CREEK Operator Protocol

Now that we have discussed the main components of a CREEK program (i.e., operators, DAGs, and streams), we delve deeper into the communication protocol between operators, this will be the abstraction level at which our run-time meta $\mu$CREEK$_r$ (Section 5) operates.

Due to the modularity of stream operators, an operator can only communicate with the operators directly connected to its in- and output ports. We call the communication protocol between two directly connected operators the "canonical stream protocol". This protocol is based on the communication protocol present in RxJava [17].

From the perspective of an operator, the protocol can be modelled as a statemachine (Section 3.2). The statemachine in Section 3.2 shows that an operator understands four distinct incoming messages (grey boxes). For each incoming message an event-handler is called. Depending on the result of the event-handler, the operator either propagates data down the stream, or ignores the event. As soon as the operator receives either an (err, e) or complete event it ends up in the final state and stops processing messages.

Source and sink actors are required to implement similar protocols, with the difference that source actors do not receive data at input ports, and sink operators do not put data on output ports. A source operator must understand one additional event in its initial state: tick. The tick event is sent by the runtime signaling the actor to propagate a value. In return it can either propagate a value ((next, v)), or it can propagate an error ((err, e)) or complete and transition to its final state.

**Compile-Time Guarantees** In the base language of CREEK, any DAG that is deployed must adhere to a set of constraints: 1) each out- and input port for each operator is connected to another port, 2) operators have exactly as many connections as ports, 3) actor sockets have exactly one in- or output port connected, and 4) no cycles are created in the DAG. If all these constraints are met the DAG can be safely deployed. If not, the program is invalid.

## 4 uCreek : Modifying DAG Structure

In Section 2 we introduced three problems with contemporary stream DSLs: non-canonical operators for non-functional concerns, mixing of functional and non-functional operators, and the inability to modify the stream execution semantics from within the DSL itself.

As part of our two-pronged approach in $\mu$Creek, we propose *structural intercession* [4] to solve the mixing of functional and non-functional operators and the fact that non-functional operators are not canonicalized across stream DSLs.

In Creek, each DAG is compiled into an internal representation, as shown in Figure 4. *structural intercession* offers hooks into this process to modify its semantics. We propose *structural intercession* because (i) it is defined separately from the domain logic and it decouples from any specific domain logic by construction, (ii) if the intercession language is expressive enough it can subsume a set of individual (non-)functional operators, and (iii) the concept of intercession is not bound to a specific stream DSL, but rather the API it exposes.

In the remainder of this section we introduce our approach to *structural intercession* in our stream DSL, $\mu$Creek$_c$, which allows the meta-level to manipulate the structure of the DAG at compile-time in a stream paradigm. We then introduce the intercession language of $\mu$Creek$_c$, and we conclude with an example of applying *structural intercession* for fusing consecutive map operators in a stream. Finally, we discuss the design principles (Section 4.3).

### 4.1 uCreekc Architecture

Before a DAG in a Creek program is deployed as a stream, the DAG is first compiled into an *internal representation* that is used as the blueprint for the stream. *structural intercession* gives the programmer control over the creation process of this internal representation at certain points. The structural intercession is exposed in the stream paradigm to the user. I..e, the compilation process is a stream itself.

**Instruction Sequences** In $\mu$Creek$_c$ each user-defined DAG is reified as an *instruction sequence* that can be transformed using a user-defined DAG. There are three different types of instructions that can be processed by this DAG. The operator and edge instruction reifiy operators and edges into their compile-time meta representation before they are added to the DAG. This allows intercession and manipulation of their definition. An operator is reified as a datastructure containing its arguments, name, and definition. An edge is reified as a datastructure containing the operator references in the DAG and the ports the edge is drawn between.

- **Operator Instruction:** For example, the instruction {:operator, :map, arg} inserts a :map operator with argument arg into the DAG.
- **Naming Instruction:** For example, the instruction {:name_it, "x"} aliases the result of the last operator instruction to the variable name "x", for reference in other instructions.

- **Edge Instruction:** For example, the instruction {:edge, "x", m, "y", n} draws an edge going from the "x" operator to the "y" operator going from port $m$ to port $n$.

Figure 3a shows an example of an instruction sequence and Figure 3b the graphical representation of the resulting internal representation.

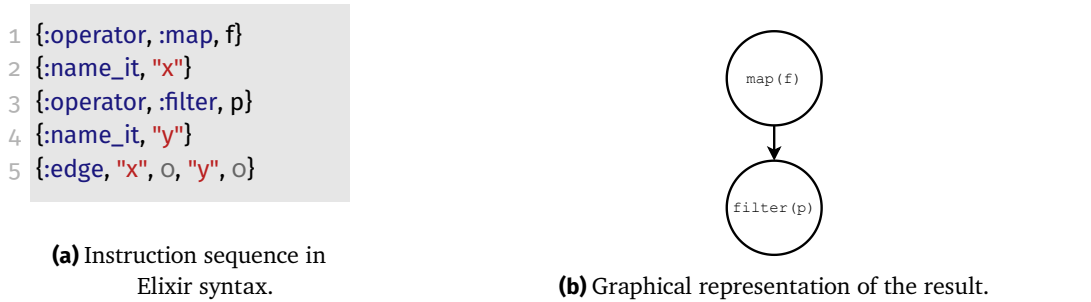

**(a)** Instruction sequence in Elixir syntax.

**(b)** Graphical representation of the result.

■ **Figure 3** Example of evaluating an instruction sequence.

**Compile-Time Meta-DAG** In $\mu\textsc{Creek}_c$ these sequences are streamed through the user-defined meta-DAG, which results in the compile-time DAG. Each instruction can be intercepted and transformed, altering the resulting internal representation. Each datum streamed is a pair of an instruction and the current internal representation. When an instruction has been applied to the internal representation by the compiler, the result – a new internal representation – is paired with the next instruction. Figure 4 depicts the pipeline of this process. First, the DAGs defined in the program are compiled to instruction sequences.

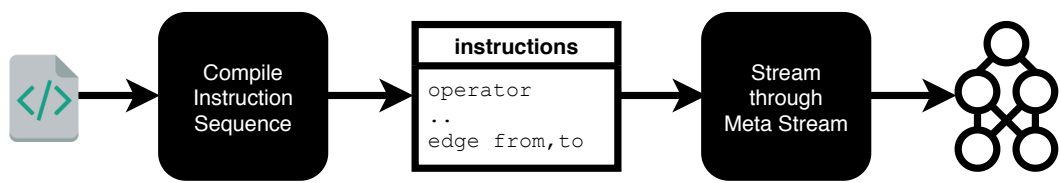

■ **Figure 4** Graphical depiction of the compiler pipeline and the *structural intercession*.

### 4.2 uCreekc Language

The goal of *structural intercession* is to influence the compilation stream of the DAG. It facilitates the manipulation of the compilation by giving control to the programmer over the instructions and DAG they define.

To this end, $\mu\textsc{Creek}_c$ offers primitive functions which manipulate operators.

- fetch!(name) fetches the operator instance bound to the name var in the DAG.
- fuse!(op1, op2) fuses exactly two existing operator instances together into a new operator. For example, if two map operators o = map(f) and p = map(g) exist in the DAG, fuse!(o,p) creates a new operator equivalent to map(fn v -> g(f(v)). Multiple operators can be fused by composing fuse calls to pairs of operators.
- swap!(op1, op2) swaps two operators in the DAG while preserving the connections.

- inputs(op) returns a list of operators that are directly connected to an input port of op.

$\mu$CREEK$_c$ also offers a set of primitives that help in modifying the partial compile-time DAG.

- add!(op) and delete!(op) adds and deletes operators, resp.
- connect!(op_from, fidx, op_to, tidx) and disconnect!(op_from, fidx, op_to, tidx) adds and deletes an edge, resp., between existing operators.

**Compile-Time Guarantees**  $\mu$CREEK$_c$ allows the programmer to transform the compilation instructions at compile-time. Because the language does not prohibit the programmer to violate the constraints mentioned in Section 4.1 an additional step is transparently executed at compile-time to ensure that the $\mu$CREEK$_c$ program did not generate an invalid DAG. If the meta-program produces an invalid DAG the program will be rejected by the compiler.

In conclusion, the $\mu$CREEK$_c$ programmer can modify the instructions with compile-time DAGs as they are streamed through the CREEK compiler. The goal of the DSL offered is to build abstractions in the form of DAGs that can be composed into specific compile-time behaviours. For example, creating a DAG that fuses map operators can be composed with a DAG that parallelizes map operators to increase the throughput of a specific stream. The DAG which is being constructed is reified as a compile-time DAG structure that can be manipulated using the $\mu$CREEK$_c$ language. The result of streaming the instruction sequence is the final internal DAG.

**Validation**  To validate our approach we implement an example of operator fusion. Operator fusion optimizes a DAG by fusing together similar operators into a single one. Semantically the result is equivalent, but the amount of operators is reduced by combining them, thus increasing performance and reducing DAG size. Operator fusion can be considered a validation because it uses all the features present in $\mu$CREEK$_c$: 1) removing and swapping operators in the DAG at compile-time, 2) inserting operators at compile-time, and 3) changing the connections between operators in the DAG based on stream events. In summary, operator fusion subsumes problems such as logging, timestamping, and parallelizing a stream. At the end of this section we briefly discuss the approach to implement these concepts in $\mu$CREEK$_c$.

Listing 7 shows the relevant code required to implement the fusion of consecutive map operators in $\mu$CREEK$_c$. The full code is shown in Listing 11 in Appendix A.

Fusing two operators together at compile-time in $\mu$CREEK$_c$ is possible because each operator is represented as a datastructure that contains the operator type (e.g., map, filter,...), and as its arguments (i.e., the function of a map operator). Fusing two map operators together involges wrapping the arguments of the individual operators into a single function that is then used as the argument of a new map operator.

To fuse operators together it suffices to focus on the edge instructions. If an edge being drawn is between two map operators $a$ and $b$ it is possible to fuse $a$ and $b$ into $c$, and connect the input of $a$ to $c$. All edges that are going to be connected to $b$ must now be connected to $c$.

The `edge` DAG defined in Listing 7 contains the transformation that will be applied to each edge in the DAG. First of all, the operators between which the edge is being put are fetched. If those two operators are not `map` operators, the instruction is unmodified (lines 17–18), otherwise it is modified as follows. To fuse the two operators (i.e., $a$ and $b$) first of all a new operator $c$ is created with the behaviour of applying $b$ after $a$. Second, $a$ and $b$ are deleted from the compile-time DAG. Third, $c$ is inserted into the compile-time DAG. Finally, the original instruction is transformed to create an edge between the output port of the source of $a$ and the input port of $c$.

All other events (i.e., operators and name events) can be passed without modification. Combining those DAGs with the `edge` DAG results in a closed DAG that can be plugged into the compiler to achieve the desired behaviour. We refer to Listing 11 in Appendix A for the full implementation.

■ **Listing 7**  Example of a meta stream fusing consecutive map operators together.

```
1  defdag edge as filter(fn event ->
2         match?({{:edge, _, _, _, _}, _, _}, event)
3     end)
4  ~> map(fn {{:edge, from, fidx, to, toidx}, dag, it} ->
5       a = fetch!(dag, from)
6       b = fetch!(dag, to)
7
8       case {a.name, b.name} do
9         {"map", "map"} ->
10            [x] = inputs(dag, a)
11            c = fuse(a, b)
12            dag = delete(dag, a)
13            dag = delete(dag, b)
14            dag = add!(dag, c)
15            {{:edge, x.ref, 0, c.ref, 0}, dag, it}
16
17         _ ->
18            {{:edge, from, fidx, to, toidx}, dag, it}
19       end
20     end)
```

To use a compile-time module in a Creek program the `structure Fusion` pragma needs to be added to the module. The compiler will use the meta-DAG defined from the Fusion module defined in Listing 7. An example of a program that uses operator fusion is shown in Listing 12 in Appendix A. Below we list additional use cases of which the implementation can be found in Appendix C.

**Timestamping**  Timestamping the values of a stream means that each datum is timestamped at the time it passes through an operator in a DAG. This can be achieved in contemporary stream DSLs by adding a `map` operator that adds a timestamp to each value. However, inserting `map` operators that wrap values with timestamps impacts every other domain logic operator that expects values without timestamps, so boxing and unboxing logic has to be added to other operators to handle the timestamps. Additionally, timestamps are stripped from the datum as soon as it passes through a regular operator.

Using $\mu\textsc{Creek}_c$, this can be implemented by making two changes to the DAG at compile-time by place a timestamping `map` operator in front of every operator in the DAG and modifying every operator that has a function as argument (e.g., `map` and `filter`) by replacing the function argument with a wrapped function that unboxes and boxes its arguments.

**Parallelization**    Parallelizing a DAG or operator requires that some operators, such as `map` operators are executed in parallel. Listing 2 shows how this is achieved in Akka Streams by duplicating operators in the DAG.

Similar to the approach for operator fusion, a `map` operator can be replaced with two operators: a `balance` operator that spreads its data over $n$ instances of the original `map` operator, and a `merge` operator which merges those parallel streams together. Just like the Akka Stream and RxJava implementation in Listing 2 this approach loses the original ordering of the data. The full implementation can be found in [23].

### 4.3 Discussion

The design of a meta-level architecture should respect certain design principles defined by earlier work in meta-level programming. We discuss each of the principles introduced by [4] and how they are addressed in $\mu\textsc{Creek}_c$

**Encapsulation**    dictates that a meta-level entity should *encapsulate* its implementation details. The meta-level programs should be written against an API that decouples the underlying implementation from the meta-level programs, enabling reusability. For $\mu\textsc{Creek}_c$ we have chosen to write meta-programs against three data structures (DAGs, edges, and operators) that offer an API that is completely decoupled from their actual implementation. The result is that $\mu\textsc{Creek}_c$ programs can be reused for implementations in other languages, as long as the meta-level architecture offers the same API.

**Stratification**    dictates that meta-level entities should be cleanly separated from base-level entities. This ensures that there as little coupling as possible between the base- and meta-level, and that the meta-level behaviour can be removed without breaking the base-level program. A meta-level program should not reference base-level entities (e.g., a base-level DAG with variable name `proceed` can not have impact on the `proceed` DAG at the meta-level), or vice versa. Additionally, a base-level operator should be prohibited to create explicit references to its meta-level representation. These references would make it impossible for the compiler to completely strip the meta-behaviour, because the base-level is coupled to the meta-level. If stratification is respected the meta-level architecture can be completely removed from an application if it is never used.

In $\mu$Creek$_c$ stratification is ensured in two ways. First of all, the meta-architecture is only loaded on-demand by using the `structure MyModule` pragma. Secondly, the meta-level does not have access to base-level concepts, and thus no references can be made.

**Structural Correspondence**   dictates that each language construct has a reified representation at the meta-level. $\mu$Creek$_c$ represent the components of a stream language at the meta-level: DAGs, edges, and operators. While strictly speaking the function arguments of operators should also be reified, we have chosen not to do so, as this requires a meta-level representation of concepts from the host language (i.e., Elixir), coupling $\mu$Creek$_c$ to a specific implementation. In conclusion, we have reified only the concepts that Creek introduces and draw the line at the host-level language.

**Unified Programming Model**   The unified programming model dictates that the paradigm for the meta-level and the base-level should be the same (e.g., streams) and has a number of advantages. On the one hand, it simplifies reasoning over the code as only a limited number of programming concepts need to be considered. On the other hand, it simplifies the interaction between both levels. For this reason, $\mu$Creek$_c$ represents the construction of the DAG as a stream of instructions. Any $\mu$Creek$_c$ program is represented as DAG, which is built out of the standard Creek operators, with the additional proceed built-in DAG.

**Portability**   The protocol (Figure 4) discussed here reifies the underlying DAG into a stream of instructions that represent changes to a DAG. These instructions can be manipulated using the base-level stream DSL and are executed by the compiler. To ensure portability this architecture cannot rely on intricacies of the underlying host language (i.e., Elixir), and therefore does not. Expressing $\mu$Creek$_c$ in any other stream language requires adapting the compiler of the stream language to generate the instruction stream and allow injecting a DAG to manipulate this stream.

## 5 uCreekr: Modifying Stream Behaviour

In Section 4 we tackled two out of three problems posited in Section 2. The third problem posited in Section 2 is that the stream execution semantics is hard-coded into the language and cannot be changed from within the language itself.

We propose – complementary to $\mu$Creek$_c$ – a *behavioural intercession* architecture. *Behavioural intercession* allows a program to modify the run-time semantics of the underlying language itself. In the context of stream DSLs this means that streams can manipulate how they are executed at run-time. For example, introduce different propagation semantics, encrypt an entire stream, or catch errors to avoid retrying a stream by intercepting the execution of event-handlers.

In the remainder of this section we introduce the architecture of *behavioural intercession* for a stream DSL by implementing it in our own stream DSL Creek (Section 5.1), discuss the intercession language of $\mu$Creek$_r$ (Section 5.2), and conclude with a validation by implementing pull-based semantics, while CREEK is push-based by default. Additionally we discuss the approach on how an entire stream can be encrypted (Section 5.3).

## 5.1 uCreekr Architecture

When a stream is executing its individual operators are exchanging messages under the hood (Section 3.2), thereby propagating the data. *Behavioural intercession* hands the programmer control over this process in order to influence its semantics.

In $\mu$CREEK$_r$ a meta-DAG is coupled to each operator in a stream. For any operator, each incoming event will be reified into a meta-event and pushed through this meta-DAG, circumventing the interpreter its default behaviour. Summarized, this means calling the appropriate event-handler and ensuring that in response the right events are emitted to the connected operators (i.e., effects).

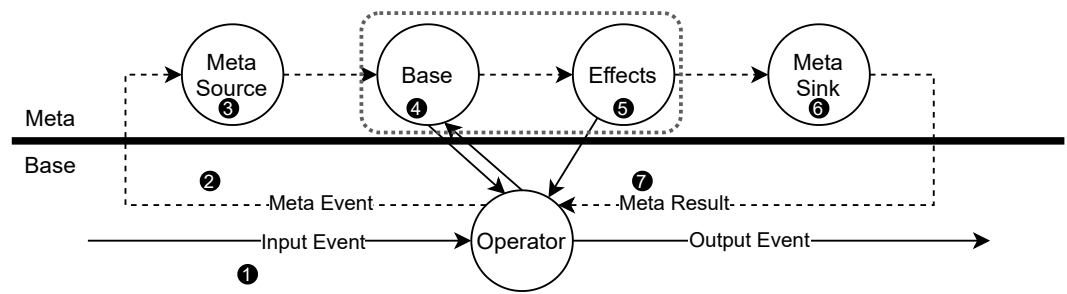

■ **Figure 5** Architecture of meta-level stream: each base-level value is reified to the meta-level, and each meta-response is deified to the base-level.

Figure 5 depicts a flow diagram of how an incoming event is processed. The process starts when an event (e.g.,(next, v)) arrives at the base-level (1). The runtime reifies the event to a meta-level representation (2), i.e., a meta-event, and emits it through its meta source actor (3). The user-defined meta stream should have two conceptual stages: base-level and effects. First of all, the stream should call the relevant base-level function for handling the meta-level event (4). Secondly, the stream should emit the necessary side-effects in response to the reply of the event-handler (5). For example, in response to (next, v), a value must be propagated to all operators connected to the output ports. The final value of the stream, representing the new operator state is captured by the sink actor (6). And finally, the result of the meta stream is installed as the new operator state (7), and the operator can receive a new base-level message.

## 5.2 uCreekr Language

In the meta-DAG operators can transform meta-level events into other meta-events, call the base-level event-handlers, and send out events to connected operators. $\mu$CREEK$_r$ offers a language that simplifies these tasks, and two open DAGs that implement default behaviour for calling the base-level and handling the side-effects. Table 2 summarizes the functions available to the operators of a meta-DAG. Three functions

◼ **Table 2** Functions available in $\mu\text{CREEK}_r$

| Function | Description |
|---|---|
| propagate_up(e) | Emit e to all the operators connected to an input port. |
| propagate_down(e) | Emit e to all the operators connected to an output port. |
| propagate_self(e) | Emit event e to itself. |
| call_base(name, arg*) | Calls the event-handler name with optional argument arg. name can be next, error, complete,tick or initialize |
| state() | Returns the current state of the operator. |

are available to handle side-effects (i.e., propagate_*), a getter function for the operator state (i.e., state), and a function that calls the base level event-handlers (i.e., call_base).

The user-defined meta-DAG has a high degree of freedom over the entire processing pipeline of an event. However, some meta-level behaviours do not require extensive changes, such as logging a stream. $\mu\text{CREEK}_r$ offers two globally defined open DAGs which implement default behaviour for both cases: proceed and effects. The former expects a meta-event as input, calls the event-handler, and returns the response of the event handler. The latter takes in the response of an event-handler and executes the effects accordingly. Combined they serve as the default meta-level behaviour.

## 5.3 Validation

To validate $\mu\text{CREEK}_r$ a form of pull-based propagation semantics is implemented in CREEK, which is push-based by default. We consider an alternate form of propagation semantics a validation because 1) related work [2, 20] posits it as one of the key problems of contemporary stream DSLs, and 2) it requires changes in the effects of sinks, sources, and operators.

The basis of pull-based semantics is a new demand message in the protocol. This message flows from the sinks to the sources whenever data must be injected into the stream.

**Source**    Listing 8 shows part of the implementation of pull-based propagation semantics in CREEK, namely the meta-DAG for a source operator. The full implementation can be found in Appendix B. In base-level CREEK, when a source receives a tick event it produces a datum. This event is sent on two occasions: when the operator is initialized, and after the operator has emitted a datum. In a pull-based system this event is omitted and will only be sent if a source demands data by sending a demand event. Additionally, a source operator must understand the demand message. In response to the demand event a source operator sends itself the tick event to trigger a datum production.

**Operator**    If an operator receives the demand message it has to propagate it upward such that it eventually reaches a source. Additionally, if the operator does not always propagate data in repsonse to an incoming event (e.g.,filter) it must propagate a new demand event upstream every time it ignores an incoming event. Operators that rely on multiple sources (e.g., zip and combineLatest [18]) can maintain some state to keep track of which sources have been sent demand and replied. This avoids sending demand twice to an operator. An example implementation can be found in Listing 14 in Appendix B.

**Sink**  First of all, as soon as the sink operator is initialized it must propagate demand upstream to initiate the data propagation. Secondly, when a sink operator receives a value event, next to the default behaviour it must also propagate a demand message upstream to initiate the propagation of a new datum. The full implementation for pull-based semantics can be found in Appendix B.

■ **Listing 8**  $\mu\text{Creek}_r$ code for source operators in a pull-based stream.

```
1  defdag tick_src as filter(&match?({p, :tick}, &1))
2          ~> base
3          ~> map(fn base_result ->f
4          case base_result do
5            {p, {state, :complete}} ->
6                effects_complete(nil, p.ds, p.us, p.pid)
7                %{p | state: state}
8            {p, {state, :tick, value}} ->
9                propagate_down({:next, value}, p.ds)
10               %{p | state: state}
11         end
12        end)
```

### 5.4  Discussion

As we did in Section 4.3 we discuss $\mu\text{Creek}_r$'s design with respect to the design principles outlined in [4].

**Encapsulation**  A $\mu\text{Creek}_r$ DAG relies on the communication protocol defined in Figure 5 and a reification of operators. This approach almost completely decouples the $\mu\text{Creek}_r$ program from the base-level, as no references can be made to the base-level. The only requirement for any language to support $\mu\text{Creek}_r$ is to reify the inter-operator communication as the canonical stream protocol and to reify the representation of base-level nodes.

**Stratification**  is achieved in $\mu\text{Creek}_r$ in the same way as it is for $\mu\text{Creek}_c$. The meta-program is only loaded if it has been explicitly imported using the "behavioural" pragma, and the meta-level does not have direct references to base-level concepts besides the values of the host language.

**Structural Correspondence**  Similar to $\mu\text{Creek}_c$ we have chosen to only represent the concepts introduced by the Creek DSL at the meta-level (i.e., operators and

■ **Listing 9** $\mu\text{Creek}_r$ code for sink operators in a pull-based stream.

```
1   defdag init_snk as filter(&match?({_, :init}, &1))
2               ~> base
3               ~> map(fn {p, {_, :ok}} ->
4                propagate_up(:demand, p.us)
5                p
6               end)
7
8   defdag next_snk as filter(&match?({_, :next, _}, &1))
9               ~> base
10              ~> effects
11              ~> map(fn {p, :ok} ->
12               propagate_up_meta(:demand, p.us)
13               p
14              end)
15
16  defdag snk_default as filter(&(not match?({_, :init}, &1)))
17              ~> filter(&(not match?({_, :next, _}, &1)))
18              ~> base
19              ~> effects
20
21  defdag sink(src, snk) do
22    src ~> dup(3) ~> (init_snk ||| snk_default ||| next_snk) ~> merge(3) ~> snk
23  end
```

events). Regular Elixir values are not reified because it would require the reification of all base-level values present in Elixir and would couple $\mu\text{Creek}_r$ to a specific host language.

**Unified Programming Model**     A $\mu\text{Creek}_r$ program is a regular Creek program, with the only difference that its domain (i.e., the data) are meta-level values. We conclude that the paradigm of $\mu\text{Creek}_r$ and Creek are the same.

**Portability**     To enable the implementation of the ideas of $\mu\text{Creek}_r$ in other languages it cannot rely on the intricacies of the underlying language. In $\mu\text{Creek}_r$, the meta-level programs only process events emitted by the runtime (i.e., reified base-level events). While not every language implements the communication protocol of Creek, we argue that the message protocol presented in Figure 5 is sufficiently high-level to implement in most languages. To port the architecture of $\mu\text{Creek}_r$ to another language the runtime must generate and emit the events from the $\mu\text{Creek}_r$ protocol, and understand the messages it can generate (Table 2). Additionally, the protocol defined in [17] was used as inspiration to the $\mu\text{Creek}_r$ protocol and is the foundation of most JVM-based stream languages [13, 19].

**Performance**   To assess the performance of our meta-level architecture, benchmarks have been run to measure it's impact on execution time. Section 6 details our benchmark setup and discusses results.

## 6   Performance Benchmarks

To evaluate the performance of the meta-level architecture we benchmarked an "identity meta" (Listing 10) and compared it to a version of Creek in which all meta-level machinery was removed, called Creek--

The identity meta (Listing 10) does not do any other computation at the meta-level than calling the base-level and executing the side-effects. This ensures that any overhead measured is introduced solely by $\mu$Creek$_r$. For performance reasons the identity meta is optimized by applying operator fusion.

To measure the performance overhead we ran two benchmarks for Creek as well as Creek--. The findings are reported below.

**Fixed Amount of Values, Varying DAG Size**   Figure 6 shows the chart for the first benchmark. In this benchmark the amount of values propagated is fixed, and the amount of operators in the DAG varies from 0 (only sink and source) to 2000.

We observe that the execution time for Creek (blue) and Creek-- (orange) are both polynomial, but that the performance of Creek is slower by a factor of 3.3. In other words, the performance of the DAG decreases polynomially with respect to the size of the DAG.

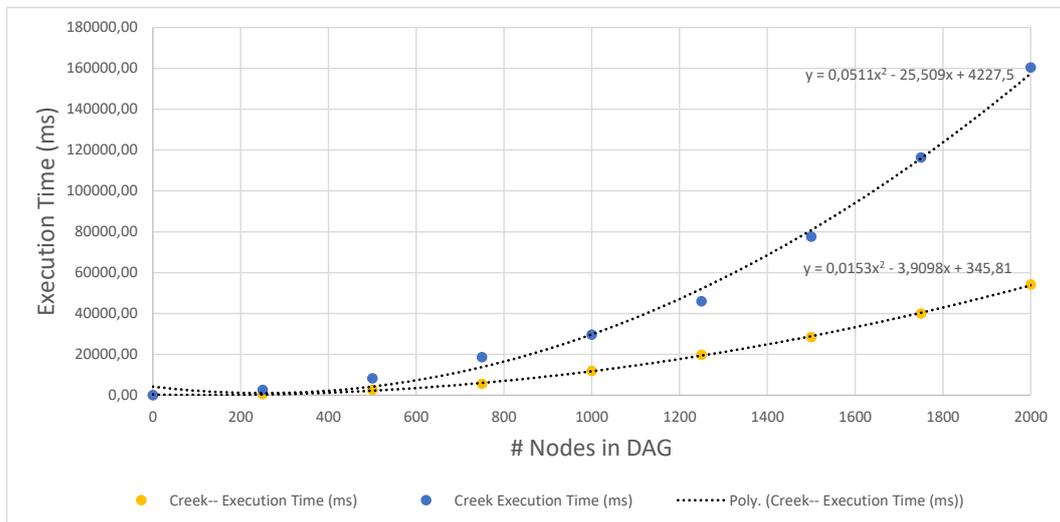

■ **Figure 6**   Execution time in miliseconds with varying DAG size and fixed load.

**Fixed DAG Size, Varying Amount of Values**   Figure 7 shows the chart for the second benchmark. In this benchmark the amount of values propagated varies from 0 to 10000, and the amount of operators in the DAG is fixed at 250.

We observe that the execution time for Creek (orange) and Creek-- (blue) are both linear with respect to the size of the input, but that the performance of Creek is slower by a factor of 77.

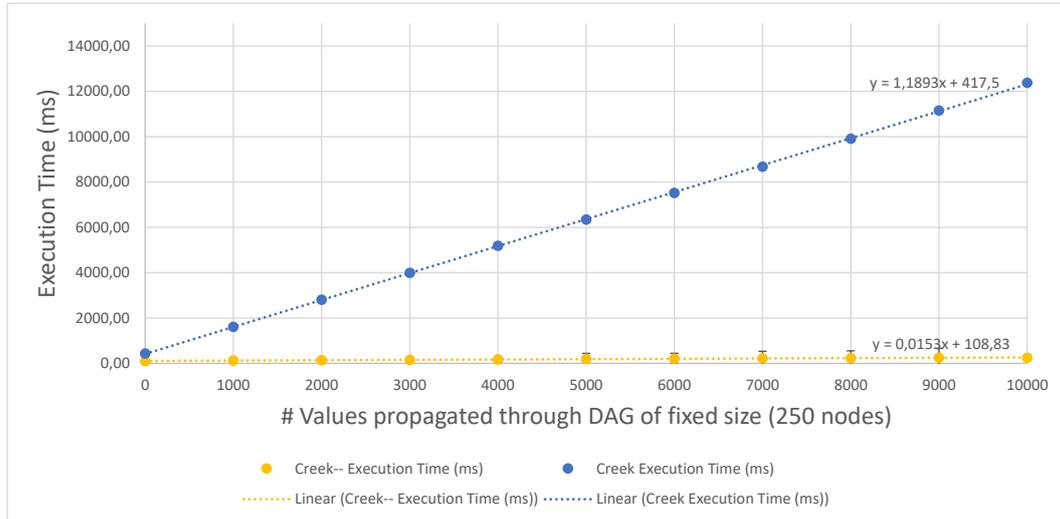

■ **Figure 7** Execution time in miliseconds with varying load and fixed DAG size.

## 7 Related Work

Streaming libraries exhibit different complexities, ranging from simple embedded DSLs such as Java Streams in Java 8, to complex pipeline architectures such as Apache Kafka. Our work focusses on the smaller DSL-based libraries such as Java Streams, RxJava, and Akka Streams. Currently, work on meta-programming in the context of stream programming is sparse. However, stream DSL programming is considered "a cousin of reactive programming" [1], so we briefly discuss the intersection of reactive programming and metaprograming.

**Metaprogramming in Reactive Programming**    In [26], Watanabe and Sawada present a reflective architecture for a functional reactive programming language called Efrp. Efrp implements a more traditional functional reactive programming approach based on time-varying values and a central clock to coordinate value propagation. An Efrp program is defined as a module containing sources, sinks and nodes depending on each other (i.e., the depedency graph). As soon as a source its value changes, all its dependencies are recomputed. The central clock ensures that glitches [5] are avoided. Recomputing a single variable and all its dependencies is called an "iteration". The reflective layer of Efrp reifies the variables in the program by name, their current value, their previous value, and the base-level expression which computes their value. A single iteration at the base-level (i.e., one variable changes) is represented as multiple iterations at the meta-level (i.e., recomputation of each of the dependencies). Because each meta-level iteration represents the update of a single dependency it is possible

■ **Listing 10** Identity meta behaviour.

```
1   defmodule IdentityMeta do
2     structure Merge
3
4     defdag operator(src, snk) do
5       src
6       ~> base()
7       ~> effects()
8       ~> snk
9     end
10
11    defdag source(src, snk) do
12      src
13      ~> base()
14      ~> effects()
15      ~> snk
16    end
17
18    defdag sink(src, snk) do
19      src
20      ~> base()
21      ~> effects()
22      ~> snk
23    end
24  end
```

to intercept the incoming value for each variable, as well as influence the final result of the variable. This approach is similar to $\mu$CREEK$_r$ and offers the same degree of control.

However, in $\mu$CREEK$_r$ the entire communication protocol can be hooked into, which is not possible in Efrp. This results in hard-coded propagation semantics, but does offer the possibility to, for example, implement encryption.

Finally, Efrp is not an embedded DSL like CREEK. This makes it easier to reify all base-level constructs all the way up to literal values. $\mu$CREEK is an embedded DSL and is therefore limited to reifying concepts that are part of the DSL.

**Propagation Semantics in Stream Libraries**    Biboudis, Palladinos, Fourtounis, and Smaragdakis present an adaptable semantics for streaming libraries [2] based on object algebras [20]. Current streaming libraries do not allow changing the propagation semantics of a streaming library and they are lacking in terms of extensibility.

The authors provide a new streaming library API which allows changing the behavior of the underlying interpreter by implementing object factories. An object factory offers an API against which a semantics can be implemented (e.g., push, pull, . . . ). Such a factory then changes the behaviour of all the built-in operators at once. The work showcases for example logging, operator fusion, push and pull factories that each provide the behavior to the operators.

The work by Biboudis, Palladinos, Fourtounis, and Smaragdakis offers increased flexibility regarding the run-time semantics and tackles a subset of the problems we discussed. The API offered in [2] relies on the work by S. Oliveira and Cook [20]. Object Algebras used in [2] rely on static type systems that at the least have generics and higher-kinded polymorphism. This means that the approach excludes a large set of programming languages, namely the dynamically typed languages without generics or higher-kinded polymorphism. Our approach, while requiring to be incorporated from the start of the design of a stream library, works on a higher abstraction level and is not tightly coupled to the underlying programming language. Furthermore, the approach offers an interface against which the semantics is to be programmed, but also requires to define an entire stream interpreter from scratch for even the smallest change. $\mu$CREEK offers the programmer a framework to express non-functional concerns, while the runtime still does most of the heavy lifting (i.e., message sending, garbage collection,...), and offers two built-in meta-level behaviours which offer default interpreter behaviour.

## 8  Conclusion

We have shown a prototypical implementation of a stream DSL embedded in Elixir called CREEK. We have extended this implementation with both meta-facilities at compile-time and run-time, and have shown by example that it is possible to implement various semantic changes from literature [2, 16] without having to change the implementation of the DSL itself. We validated our approach by addressing the three problems listed in Section 2 and demonstrated the applicability of the protocol. All these use-cases showcase an increase in extensibility of the framework, and flexibility for the user of the framework to tune it to its expectations.

**Future Work**  We argued in this text that our meta-level is expressive enough to move the meta-concerns such as error handling, buffering, and propagation semantics from the base-level code to the meta-level. Additionally, we have argued that compile-time intercession is expressive enough to deal with meta-concerns such as operator-fusion and DAG optimization. However, the composability of the run-time meta-level is lacking. In future work it is worth investigating how multiple run-time meta-DAGs can be composed safely and efficiently to form a single behavior. Consider the scenario where a stream should not only be pull-based, but also timestamped. Another path worth investigating is extending the intercession to pairs of operators by means of adapting the behavior of ~>. This operator is in charge of making the connection between two or more operators and has the potential to change the behavior of how the operators communicate. Consider the scenario where one wants to connect to a remote stream on a sensor device. Hard- and software failure is very common in IoT systems, so the software should be able to deal with this in an elegant way. We propose to override the connect operator to add additional monitoring mechanics such as leasing [3]. Additionally, we would like to see support for higher-order operators. We argue that the typical scenario of "one value in, possibly one value out" does not apply for these operators, because a single input value (a higher order stream) can cause the emission of a possibly infinite amount of values. However, higher-order operators rewrite the DAG at run-time by adding sources to operators at run-time. Assuming an operator is executing with a meta behaviour $M$ and a new operator is added as its source—as is the case in flatMap—the newly added source should understand the same meta-level messages as the original operator.

Finally, the composition of operators happens without any compile-time checks of the type of values. Possible avenues to improve upon this are contracts for run-time verification, or type annotations for compile-time verification of DAGs.

**Acknowledgements**    This work has been funded by the Fonds Wetenschappelijk Onderzoek project D3-CPS.

**A**   **Operator Fusion**

■ **Listing 11**   Example of a meta-stream fusing consecutive map operators together.

```
1  defmodule Fusion do
2    use Structural
3
4    defdag metadag(src, snk) do
5      src
6      ~> dup(3)
7      ~> (edge ||| default_operators ||| default_name )
8      ~> merge(3)
9      ~> proceed
10     ~> snk
11   end
12
13   defdag edge as filter(fn event ->
14               match?({{:edge, _, _, _, _}, _, _}, event)
15             end)
16           ~> map(fn {{:edge, from, fidx, to, toidx}, dag, it} ->
17             a = fetch!(dag, from)
18             b = fetch!(dag, to)
19             case {a.name, b.name} do
20               {"map", "map"} ->
21                 [x] = inputs(dag, a)
22                 c = fuse(a, b)
23                 dag = delete(dag, a)
24                 dag = delete(dag, b)
25                 dag = add!(dag, c)
26                 {{:edge, x.ref, 0, c.ref, 0}, dag, it}
27               _ ->
28                 {{:edge, from, fidx, to, toidx}, dag, it}
29             end
30           end)
31
32   defdag others as filter(fn event ->
33               not match?({{:edge, _, _, _, _}, _, _}, event)
34             end)
35 end
```

■ **Listing 12** CREEK program which enables operator fusion for a DAG computing the square of a number plus 1.

```
1  defmodule MyFusedDAGs do
2    use Creek
3    structure Fusion
4
5    defdag test(src, snk) do
6      src
7      ~> map(fn x -> x * x end)
8      ~> map(fn x -> x + 1 end)
9      ~> snk
10   end
11
12   def main() do
13     source = Creek.Source.list([1, 2, 3, 4])
14     sink  = Creek.Sink.doForAll(fn x -> IO.puts x end)
15     Creek.Runtime.deploy(test(), [src: source, snk: sink])
16   end
17  end
```

## B  Pull-Based Semantics

This section contains the source code for pull-based semantics in Creek. The first version in Listing 13 is a naive approach that makes sinks send a `demand` message every time they receive a value. While this approach works, it can be fine-tuned to remove some duplicate `demand` messages. This implementation is shown in Listing 14. In that implementation a `demand` message is sent only to operators that have delivered upon their previous demand. That is to say, demand will only be sent if the previous demand was met with a reply from the upstream.

■ **Listing 13**  Complete Implementation of Pull-Based Propagation Semantics in $\mu\text{Creek}_r$.

```
1  defmodule Pull do
2    use Behaviour
3
4    # This DAG handles all events with default behaviour.
5    defdag default as base ~> effects
6
7    ############################################################
8    # Operators
9
10   # This DAG propagates every demand message upstream.
11
12   defdag forward_demand as filter(&match?({_, :demand}, &1))
13                    ~> map(fn {p, :demand} ->
14                        propagate_upstream(:demand, p.us)
15                        {p, :ok}
16                      end)
17
18   # If an operator does not propagate a vlaue the demand is "lost".
19   # As soon as no value is propagated in response, a new demand is sent.
20
21   defdag opr_next as filter(&match?({_, :next, _}, &1))
22                  ~> base()
23                  ~> map(fn {p, base_response} ->
24                      if match?({_, :skip}, base_response) do
25                        propagate_upstream(:demand, p.us)
26                      end
27                      {p, base_response}
28                    end)
29                  ~> effects()
30
31   defdag opr_default as filter(&(not match?({_, :demand}, &1)))
32                    ~> filter(&(not match?({_, :next, _}, &1)))
33                    ~> default
34
35   defdag operator(src, snk) do
36     src
```

```
37    ~> dup(3)
38    ~> (opr_default ||| forward_demand ||| opr_next)
39    ~> merge(3)
40  end
41    ~> snk
42
43
44  ################################################################
45  # Sources
46  # Intercept the init event for sources to stop them from ticking themselves.
47  defdag init as filter(&match?({_, :init}, &1))
48            ~> base()
49            ~> map(fn p, {state, :initialized}} ->
50                # Here we would normally send tick to ourselves,
51                # but we dont (pull).
52                {%{p | state: state}, :ok}
53              end)
54
55  # This DAG handles the demand messages.
56  defdag demand_src as filter(&match?({p, :demand}, &1))
57              ~> map(fn {p, :demand} ->
58                # If a source receives demand it ticks itself.
59                propagate_self({:tick}, p.pid)
60                {p, :ok}
61              end)
62
63  # If a source gets a tick event (from itself)
64  # it will produce a value and tick itself again.
65  # We intercept that tick and stop from sending it.
66  defdag tick_src as filter(&match?({p, :tick}, &1))
67            ~> base()
68            ~> map(fn base_result ->
69              case base_result do
70                {p, {state, :complete}} ->
71                  effects_complete(nil, p.ds, p.us, p.pid)
72                  {%{p | state: state}, :ok}
73                {p, {state, :tick, value}} ->
74                  propagate_downstream({:next, value}, p.ds)
75                  {%{p | state: state}, :ok}
76              end
77            end)
78
79  # This DAG handles all events except the ones we intercepted.
80  defdag src_default as filter(&(not match?({_, :init}, &1)))
81              ~> filter(&(not match?({_, :meta, _}, &1)))
82              ~> filter(&(not match?({p, :tick}, &1)))
83              ~> default
84
85  defdag source(src, snk) do
```

```
 86   src
 87   ~> dup(4)
 88   ~> (src_default ||| init ||| demand_src ||| tick_src)
 89   ~> merge(4)
 90   ~> snk
 91 end
 92
 93 ################################################################
 94 # Sinks
 95
 96 # When a sink is initialized it normally doesnt do anything.
 97 # In pull-based we must send the first pull message.
 98 defdag init_snk as filter(&match?({_, :init_snk}, &1))
 99               ~> base()
100               ~> map(fn {p, {_, :ok}} ->
101                   # Normally no side-effects happen in a sink init,
102                   # but now e must propagate demand upstream.
103                   propagate_upstream(:demand, p.us)
104                   {p, :ok}
105                 end)
106
107 # This DAG ensures that a new demand is sent when a next value arrived.
108 defdag next_snk as filter(&match?({_, :next, _}, &1))
109               ~> default()
110               ~> map(fn {p, :ok} ->
111                   # After the default, we send demand upstream.
112                   propagate_upstream(:demand, p.us)
113                   {p, :ok}
114                 end)
115
116 # This DAG handles all events except the ones we intercepted.
117 defdag snk_default as filter(&(not match?({_, :init_snk}, &1)))
118               ~> filter(&(not match?({_, :next, _}, &1)))
119               ~> default()
120
121 defdag sink(src, snk) do
122   src
123   ~> dup(3)
124   ~> (init_snk ||| snk_default ||| next_snk)
125   ~> merge(3)
126   ~> snk
127 end
128 end
```

■ **Listing 14**  Implementation of pull-based sematics that works for two-pronged operators such as zip and merge.

```
1   defmodule SmartPull do
2     use Behaviour
3
4     defdag default as base ~> effects
5
6     # Operators
7     defdag forward_demand as filter(&match?({_, :meta, :demand, from}, &1))
8                        ~> map(fn {p, :meta, :demand, from} ->
9                             demanded = p.meta_state
10                            to_demand = p.us |> Enum.filter(&(not MapSet.member?(demanded, &1)))
11                            propagate_upstream_meta(:demand, to_demand, p.pid)
12                            meta_state = MapSet.new(p.us)
13                            {%{p | meta_state: meta_state}, :ok}
14                          end)
15
16    defdag opr_next as filter(&match?({_, :next, _, _}, &1))
17                   ~> map(fn {p, :next, v, from} ->
18                        meta_state = p.meta_state |> MapSet.delete(from)
19                        {%{p | meta_state: meta_state}, :next, v, from}
20                      end)
21                   ~> base()
22                   ~> map(fn {p, base_response} ->
23                        if match?({_, :skip}, base_response) do
24                          demanded = p.meta_state
25                          to_demand = p.us |> Enum.filter(&(not MapSet.member?(demanded, &1)))
26                          propagate_upstream_meta(:demand, to_demand, p.pid)
27                        end
28                        {p, base_response}
29                      end)
30                   ~> effects()
31
32    defdag opr_default as filter(&(not match?({_, :meta, :demand, _}, &1)))
33                      ~> filter(&(not match?({_, :next, _, _}, &1)))
34                      ~> filter(&(not match?({_, :init_opr}, &1)))
35                      ~> default
36
37    defdag init_opr as filter(&match?({_, :init_opr}, &1))
38                    ~> base()
39                    ~> map(fn {p, resp} ->
40                         p = %{p | meta_state: MapSet.new()}
41                         {p, resp}
42                       end)
43                    ~> effects()
44
45    defdag operator(src, snk) do
46      src
```

```
47  ~> dup(4)
48  ~> (opr_default ||| forward_demand ||| opr_next ||| init_opr)
49  ~> merge(4)
50  ~> snk
51  end
52
53  # Sources
54
55  defdag init_src as filter(&match?({_, :init_src}, &1))
56                ~> base()
57                ~> map(fn {p, {state, :initialized}} ->
58                  {%{p | state: state}, :ok}
59                  end)
60
61  defdag demand_src as filter(&match?({p, :meta, :demand, _}, &1))
62                ~> map(fn p, :meta, :demand, _} ->
63                  # If a source receives demand it ticks itself.
64                  send_self({:tick}, p.pid)
65                  {p, :ok}
66                  end)
67
68  defdag tick_src as filter(&match?({p, :tick}, &1))
69                ~> base()
70                ~> map(fn base_result ->
71                  case base_result do
72                  {p, {state, :complete}} ->
73                    effects_complete(nil, p.ds, p.us, p.pid)
74                    {%{p | state: state}, :ok}
75                  {p, {state, :tick, value}} ->
76                    propagate_downstream({:next, value}, p.ds, p.pid)
77                    {%{p | state: state}, :ok}
78                  end
79                  end)
80
81  defdag src_default as filter(&(not match?({_, :init_src}, &1)))
82                ~> filter(&(not match?({_, :meta, _, _}, &1)))
83                ~> filter(&(not match?({p, :tick}, &1)))
84                ~> default
85
86  defdag source(src, snk) do
87   src
88   ~> dup(4)
89   ~> (src_default ||| init_src ||| demand_src ||| tick_src)
90   ~> merge(4)
91   ~> snk
92  end
93
94  # Sinks
```

```
95   defdag init_snk as filter(&match?({_, :init_snk}, &1))
96              ~> base()
97              ~> map(fn {p, {_, :ok}} ->
98                 propagate_upstream_meta(:demand, p.us, p.pid)
99                 {p, :ok}
100                end)
101
102  defdag next_snk as filter(&match?({_, :next, _, _}, &1))
103              ~> default()
104              ~> map(fn {p, :ok} ->
105                 propagate_upstream_meta(:demand, p.us, p.pid)
106                 {p, :ok}
107                end)
108
109  defdag snk_default as filter(&(not match?({_, :init_snk}, &1)))
110              ~> filter(&(not match?({_, :next, _, _}, &1)))
111              ~> default()
112
113  defdag sink(src, snk) do
114    src
115    ~> dup(3)
116    ~> (init_snk ||| snk_default ||| next_snk)
117    ~> merge(3)
118    ~> snk
119  end
120  end
```

### C    Additional Use Cases

To showcase $\mu\textsc{Creek}_r$ and $\mu\textsc{Creek}_c$ we have implemented additional use cases using both the run-time and compile-time meta-levels. The code for each of these is contained in [23], and is not listed here. To make navigation in the source code easier we include a table that maps the uses cases on their respective implementation.

### C.1  Compile-Time Use Cases

**Operator Fusion**    Operator fusion "fuses" together consecutive map and filter operators into a single operator.

| | |
|---|---|
| Meta-Level | creek/lib/lang/compiler/meta_examples/merge.ex |
| Base-Level | creek/lib/examples/example_ctm_merge_merge.ex |

**Parallelization**    The parallelization compile-time meta allows the programmer to mark certain map operators with the "parallel: $n$" option. The compiler will create $n$ parallel instances of the given map operator at compile-time.

| | |
|---|---|
| Meta-Level | creek/lib/lang/compiler/meta_examples/par.ex |
| Base-Level | creek/lib/examples/example_ctm_par.ex |

### C.2  Run-Time Use Cases

**Encryption**    The underlying concept of encrypting a stream is that values need to "unboxed" before they are processed by an operator (i.e., removing the encryption) and "boxing" them after calling the base-level (i.e., encrypting the result). This needs to be done at every operator in the stream, assuming the source produces encrypted values.

| | |
|---|---|
| Meta-Level | creek/lib/lang/runtime/meta/encrypted.ex |
| Base-Level | creek/lib/examples/example_rtm_encrypted.ex |

**Logging**    Logging is can be used to debug streams, or to log the operations of a program at deployment. The given example logs every value propagating through the stream from the source to the sink.

| | |
|---|---|
| Meta-Level | creek/lib/lang/runtime/meta/logging.ex |
| Base-Level | creek/lib/examples/example_rtm_logging.ex |

**Pull Semantics**    Pull-semantics change the way data is propagated through the stream. Unchanged, $\textsc{Creek}$ pushes values through the stream. That is to say, the sources dictate the rate at which data is propagated. In a pull-based scenario the sinks dictate the rate at which data is propagated through the stream.

| | |
|---|---|
| Meta-Level | creek/lib/lang/runtime/meta/pull.ex |
| Base-Level | creek/lib/examples/example_rtm_pull.ex |

**Smart Pull Semantics**   This variation of the previousely mentioned pull-based semantics removes some redundant pull messages. The redundant messages are caused by operators that have two sources. If one source produces a value the operator will request data from both sources in the previous example. In this version the operator keeps track of which source has been "demanded" and which has not.

| | |
|---|---|
| Meta-Level | creek/lib/lang/runtime/meta/pullsmart.ex |
| Base-Level | creek/lib/examples/example_rtm_pull.ex |

## About the authors

**Christophe De Troyer** is a Ph.D candidate at the Software Languages Lab (SOFT) of the Vrije Universiteit Brussel in Belgium. His research is centered around distributed reactive programming, and more specifically stream-based programming. Contact him at cdetroyer@vub.ac.be

**Jens Nicolay** Jens Nicolay is a Professor at the Software Languages Lab (SOFT) of the Vrije Universiteit Brussel in Belgium. His doctoral research and work as a professor has mainly focused on program analysis with security as an important application. Contact him at jens.nicolay@vub.be.

**Wolfgang De Meuter** is a professor at the Software Languages Lab (SOFT) of the Vrije Universiteit Brussel in Belgium. His current research is mainly situated in the field of distributed programming, concurrent programming, reactive programming, and big data processing. His research methodology varies from more theoretical approaches (e.g., type systems) to building practical frameworks and tools (e.g., crowd-sourcing systems). Contact him at Wolfgang. De.Meuter@vub.be.